\documentclass[submission,copyright,creativecommons]{eptcs}
\providecommand{\event}{SR 2014}

\usepackage{amsmath}
\usepackage{amssymb}
\usepackage{amsthm}
\usepackage{tikz}

\newcommand{\takeout}[1]{}
\newcommand{\calG}{\mathcal{G}}
\newcommand{\calA}{\mathcal{A}}
\newcommand{\calB}{\mathcal{B}}
\newcommand{\calC}{\mathcal{C}}

\newcommand{\calK}{\mathcal{K}}

\newtheorem{theorem}{Theorem}

\newtheorem{lemma}{Lemma}[section]

\theoremstyle{definition}

\title{Games with recurring certainty\thanks{This work was partly supported by European project Cassting (FP7-ICT-601148).}}

\author{Dietmar Berwanger
\institute{Laboratoire Sp\'ecification et V\'erification\\CNRS \& ENS Cachan, France}
\email{dwb@lsv.ens-cachan.fr}
\and
Anup Basil Mathew
\institute{Institute of Mathematical Sciences\\
Chennai, India}
\email{anupbasil@imsc.res.in}
}
\def\titlerunning{Games with Recurring Certainty}
\def\authorrunning{D. Berwanger \& A. B. Mathew}
\begin{document}
\maketitle

\begin{abstract}
Infinite games where several players seek to coordinate under imperfect
information are known to be intractable, unless the information flow
is severely restricted. Examples of undecidable cases typically feature 
a situation where players become uncertain about
the current state of the game, and this uncertainty lasts forever.

Here we consider games where the players attain certainty about the 
current state over and over again along any play.
For finite-state games, we note that this kind of \emph{recurring} certainty
implies a stronger condition of \emph{periodic} certainty, that is,
the events of state certainty ultimately occur at uniform, regular intervals.
We show that it is decidable whether a given game presents recurring certainty,
and that, if so, the problem of synthesising coordination strategies 
under $\omega$-regular winning conditions is solvable.
\end{abstract}

\section{Introduction}

Automated synthesis of systems that are correct by construction 
is a persistent ambition of computational engineering. 
One major challenge consists in controlling 
components that have only partial information about the 
global system state. 
Building on automata and game-theoretic foundations,
significant progress has been made
towards synthesising finite-state components that interact 
with an uncontrollable environment either individually, 
or in coordination with other controllable components --- 
provided the information they 
have about the global system is distributed hierarchically~\cite{PnueliRos89,KupfermanVar01,Kai06}.
Absent such restrictions, however, the problem of coordinating
two or more components of a distributed system with 
non-terminating executions is generally 
undecidable~\cite{PnueliRos90,ArnoldWal07}. 

The distributed synthesis problem can be formulated alternatively
in terms of games between
$n$~players (the components) that move along the edges of a finite graph 
(the state transitions of the global system) with imperfect information about the the current position and the moves of the other players. 
The outcome of a play 
is an infinite path (system execution) 
determined by the joint actions of the players 
and moves of Nature (the uncontrollable environment). 
The players have a common winning condition:
that the play corresponds to a correct execution with respect to 
the system specification, no matter how Nature moves.
Thus, distributed synthesis under partial information corresponds to 
the 
problem  of constructing a winning profile of finite-state 
strategies 
in a coordination game with imperfect information, 
which was shown to be undecidable already in \cite{Reif84}, 
for the basic setting of two players with a reachability condition, 
and in~\cite{Janin07}, for more complex winning conditions.

The cited undecidability arguments share a basic scenario: 
two players -- he and she -- become uncertain 
about the current state of the game, due to moves of Nature.
As her (partial) knowledge of the state differs from his, 
and their actions need to respect the uncertainty of both, she
needs to keep track not only of what she or he
knows about the game state, 
but also, e.g., of what he knows about 
what she knows that he knows, and so on.
The scenario, set up so that the 
uncertainty never vanishes, leads to undecidability as  
the knowledge hierarchies grow unboundedly while the play proceeds 
\cite{BerwangerKai10}.
  
The \emph{information fork} criterion of \cite{FinkbeinerSch05} 
identifies distributed system architectures that may allow the knowledge of 
players to develop differently, for an unbounded number of rounds. 
Nevertheless, information forks may not cause undecidability in every context, for instance, if the ``forked knowledge'' is irrelevant for enforcing the 
winning condition, or if the effect of forking can be undone 
within a few rounds every time it occurs.

In this paper, we 
consider $n$-player games with imperfect information 
where the uncertainty of players about the 
game state cannot last forever. 
Our intuition of \emph{recurring certainty} is that, whenever players are uncertain about the state of 
the game during a play, it takes only finitely many 
rounds until they can deduce the current state with certainty, and 
it becomes common knowledge among them. A faithful formalisation of this common knowledge property would most likely be undecidable. 
Thus, we resort to a weakening which intuitively states that 
the current state is evident to all players.

We show that the following two questions are decidable:
\begin{enumerate}
\item[$\bullet$] Given an $n$-player game structure with imperfect information, 
does it satisfy the condition of recurring certainty?
\item[$\bullet$] Given a game with recurring certainty and an $\omega$-regular  winning condition, does the grand coalition have a winning strategy?
\end{enumerate}

Towards this, we first prove that, under recurring certainty, 
the intervals where the current state of the game is not common knowledge are 
bounded uniformly. We call this periodic certainty. 
Then, we show that  the perfect-information \textit{tracking}~\cite{BKP11} of a
 game with periodic certainty is finite. This allows to solve the synthesis problem.

\textit{Acknowledgement}. The authors thank Marie Van den Bogaard for useful discussions on related topics and for proof-reading this paper.

\section{Coordination games with imperfect information}

Our game model is close to that of concurrent games \cite{AlurHK02}.
There are $n$ players $1$, \dots, $n$ and a distinguished agent 
called nature. The \emph{grand coalition} is the set $N = \{1, \dots, n \}$ 
of all players. 
We refer to a list of elements
$x=(x^i)_{i\in N}$, one for each player, as a \emph{profile}.  

For each player~$i$ we fix a set $A^i$ of \emph{actions} and a set $B^i$ of observations, 
finite unless stated otherwise.
The \emph{action space} $A$ consists of all action profiles.
A \emph{game structure} $G = (V, \Delta, (\beta^i)_{i \in N})$ 
consists of a finite set $V$ of \emph{states}, a relation $\Delta \in V \times A \times V$ of simultaneous \emph{moves} labelled by action profiles, 
and a profile of \emph{observation} functions $\beta^i: V \to B^i$.
We assume that each
state has at least one outgoing move for every
action profile, i.e., $\Delta(v,a) \neq
\emptyset$, for all $v\in V$ and all $a \in A$.

Plays start at an initial state $v_0 \in V$ known to all players, and proceed in rounds. 
In a round, all players~$i$ choose an action 
$a^i \in A^i$ simultaneously, then  
nature chooses a successor state $v' \in \Delta(v, a)$ and
each player~$i$ receives the observation $\beta^i( v')$.  
Notice that the players are not directly informed about the action
chosen by other players nor the state chosen by nature. However, 
we assume that the player's own action is part of his observation at the 
target state.  
Formally, a \emph{play} 
is an infinite sequence $\pi = v_0, a_0, v_1, a_1, \dots$ 
alternating between positions and action profiles 
with $(v_{\ell}, a, v_{\ell + 1}) \in \Delta$, for all $\ell \ge 0$.
A \emph{history} is a prefix $v_0, a_0, \dots, a_{\ell-1},
v_{\ell}$ of a play. 
The observation function extends from states 
to histories and plays $\pi = v_0, a_0, v_1, a_1, \dots$ 
as $\beta^i( \pi ) = \beta^i( v_0 ), \beta^i (v_1), \dots$.
We say that 
two histories $\pi, \pi'$ are \emph{indistinguishable} to Player~$i$, and write 
$\pi \sim^i \pi'$, if $\beta^i(\pi) = \beta^i(\pi')$. 
This is an equivalence relation, and its classes are 
called the \emph{information sets} of Player~$i$.

A \emph{strategy} for Player~$i$ is a mapping $s^i: (V\kern-2pt A)^*V \to A^i$
from histories to actions
such that  $s^i ( \pi ) = s^i( \pi')$, for any pair ~$\pi \sim^i \pi'$ 
of indistinguishable histories.
We denote the set of all strategies of Player~$i$ 
with $S^i$ and the set of all strategy profiles by $S$.
A history or play $\pi = v_0, a_0, v_1, a_1, \dots$ 
\emph{follows} the strategy $s^i \in S^i$, if  
$a_{\ell}^i = s^i( v_0, a_0, v_1, \dots, a_{\ell-1}, v_\ell)$ for every  $\ell > 0$.
For the grand coalition, the play $\pi$ follows a strategy profile
$s$, 
if it follows all strategies $s^i$.
The set of  possible \emph{outcomes} of a strategy profile~$s$ 
is the set of plays that follow~$s$.

A \emph{winning condition} over a game structure $G$ is a set 
$W \subseteq (V\kern-2pt A)^\omega$ of plays. A \emph{game} $\calG = (G, W)$ 
consists of a game structure and a winning condition. 
We say that a play $\pi$ on $G$ is winning in $\calG$ if $\pi \in W$; 
a strategy profile $s$ is winning in $\calG$, 
if all its possible outcomes are so. 
To describe winning conditions,
we use a colouring function $\gamma: V \to C$ with a finite range $C$ of colours, and refer to the set $W \subseteq C^\omega$ 
of all plays $v_0, a_0, v_1,
a_1, \dots$ with $\gamma( v_0 ), \gamma( v_1), \dots  \in W$.
In this paper, we assume that the colouring is \emph{observable} to each player~$i$, 
that is, $\beta^i( v ) \neq
\beta^i( v')$ whenever $\gamma( v ) \neq \gamma( v')$.

We consider coordination games over finite game
structures where the winning condition is given by finite-state
automata. (See \cite{GraedelTW02}, for a comprehensive background.)
Given such a game $\calG$, 
we are interested in the following questions: 
(1) Does the grand coalition have a winning strategy
  profile in $\calG$? and
(2) How to synthesise (distributed) winning strategies, if they exists?

\section{Recurring certainty}
\label{rec_cer}

We consider a class of games where the uncertainty of players 
about the current state is temporary 
and vanishes after a finite number of rounds. 

To explain our notion of certainty, 
we introduce a fictitious player, 
let us call him Player $0$, 
who is less informed than any actual player. 
He does not contribute to joint actions (i.e., his action set $A^0$ is a singleton), and 
his observation function is a coarsening of all observations of other 
players: for any pair $v, v'$ of game states, $\beta^0( v ) = \beta^0 (v')$
whenever $\beta^i (v) = \beta^i( v')$ for some player $i$. 
Thus, for histories $\pi,\pi'$, we have $\pi \sim^0 \pi'$, 
whenever $\pi \sim^i \pi'$ for some player $i$ 
(the converse does not hold, in general). 
  
For a given game structure $G$, we say that the grand coalition
\emph{attains certainty} 
at history $\pi = v_0, a_0, \dots, a_{\ell-1}, v_{\ell}$, if any indistinguishable history $\pi'\sim^0\pi$ 
ends at the same state $v_{\ell}$.
An infinite play $\pi$ has \emph{recurring certainty}, if the grand coalition attains certainty at infinitely many of its histories. 
Finally, we say that the game structure $G$ has recurring certainty, 
if this is the case for every play in $G$. 

As a simple example of a game with recurring certainty, 
consider the infinite repetition of 
a finite extensive game with imperfect information
 where the root is a perfect-information node, i.e., 
it is distinguishable from any 
other node, for every player. 
Likewise, games on graphs with the property that every cycle passes 
through a perfect-information state have recurring certainty.

We will also encounter the following stronger property. 
A game structure $G$ has 
\emph{periodic certainty} if there exists a uniform bound $t \in \mathbb{N}$ 
such that for every play $\pi$ in $G$, every history $\rho$ of $\pi$ 
has a continuation $\rho'$ by at most $t$ rounds in $\pi$, 
such that the grand coalition attains certainty at $\rho'$.

\subsection{Recognising games with recurring certainty}

Our first result states that 
recurring certainty is a regular property of plays in finite game structures. 

\begin{lemma}
For any finite game structure, the set of plays where the grand coalition has 
recurring certainty is recognisable by a finite-state automaton.   
\end{lemma}

\begin{proof}
Let us fix a finite game structure $G$. 
First, we construct a word automaton~$\calA$ over the alphabet $AV$
that recognises histories $\rho$ at 
which the grand coalition does not attain certainty. To witness this, 
the automaton guesses a second history $\rho'$ (of the same length) 
that is $\sim^0$-indistinguishable from $\rho$ and ends at a different state.

The state space of $\calA$ consists of pairs of game states in $V$, 
plus a sink. The first component of the automaton state keeps track of the 
input history and the second one of the uncertainty witness 
that is guessed nondeterministically. The transition function 
ensures that both components evolve according to the moves
available in the game structure and 
yield the same observation to all players; otherwise, they lead to the sink. 
Accepting states are those where the first and the second component differ.

By complementing the automaton $\calA$, we obtain an automaton $\overline{\calA}$ 
that accepts the set of histories at which the grand coalition
attains certainty (plus sequences that do not correspond to histories, 
which can be excluded easily by intersection with the unravelling of $G$). 
Next, we determinise $\overline{\calA}$ and view 
the outcome as a deterministic B\"uchi automaton $\calB$
which accepts the input word, if it hits the set of final states
infinitely often.  
Thus, $\calB$ accepts all plays where the grand coalition has recurring certainty. 
\end{proof}

The synchronous 
product of the deterministic B\"uchi automaton $\calB$ constructed above 
with the game structure $G$ is universal, i.e accepts every play of $G$, if and only if, 
$G$ has recurring certainty. 

\begin{theorem}
The question whether a given game structure has 
recurring certainty is decidable.
\end{theorem} 
 
A further consequence of the automaton construction is 
that we obtain a uniform bound on the distance between two rounds 
at which the grand coalition attains certainty.    

\begin{theorem}
Every game with recurring certainty also has periodic certainty.
\end{theorem}
 
\begin{proof}
Let $G$ be a game structure with recurring certainty, $\calB$ the deterministic B\"uchi automaton constructed for~$G$ as above, and let $t$ be the number of states in $\calB$ plus one.
Towards a contradiction, suppose there exists a play $\pi$ in $G$ 
with a collection of $t > |\calB|$ many 
consecutive histories $\rho_0, \rho_1, \dots \rho_t$ 
at which the grand coalition does not attain certainty. 
Accordingly, the uniquely determined run of $\calB$ on input $\pi$ hits
no accepting state of the automaton $\calB$ while reading the continuation of 
$\rho_0$ up to $\rho_t$. On the other hand, as $t > |\calB|$, 
there exists a state in $\calB$
that is reached by two different histories, 
say $\rho_k$ and $\rho_\ell$, with $0 \le k \le \ell \le t$. 
Now we consider the play
$\pi'$ on $G$ that begins with
$\rho_\ell$, and then repeats the continuation of $\rho_k$ up to
$\rho_\ell$ forever. Thus, the run of $\calC$ on $\pi'$ will finally not
hit any accepting state and be rejected, in contradiction to our
assumption that $G$ has recurring certainty.
\end{proof}

\subsection{Winner determination and strategy synthesis}

\begin{theorem}
Let $\calG$ be a coordination game with an $\omega$-regular winning condition.
If $\calG$ has recurring certainty, then the question whether the grand
coalition has a winning strategy profile is decidable and the strategy synthesis problem is effectively 
solvable.
\end{theorem}

Our argument relies on the tracking construction proposed in~\cite{BKP11}
that eliminates 
imperfect information in $n$-player games by an unravelling process 
that generates epistemic models of the player's information along the stages of a play. 
An \emph{epistemic model} for a game structure $\calG$ is a 
Kripke structure $\calK = (K, (Q_v)_{v \in V}, (\sim^i)_{i \in N})$ 
over a set $K$ of histories in $\calG$, equipped with 
predicates $Q_v$ designating the histories that end in state 
$v \in V$ and the players' indistinguishability relations $\sim^i$.
The construction keeps track of how the knowledge of players is updated
by generating, for each epistemic model $\calK$, a set of 
successor models along
tuples $(a_k)_{k \in K}$ of action profiles $a_k \in A$
compatible with the player's current knowledge, i.e.
for every $i \in N$ and for all $k, k' \in K$ with
$k \sim^i k'$, we have $a_k^i = a_{k'}^i$. 
This leads to a possibly disconnected epistemic model with universe 
$K' = \{k a_k v \mid k \in K, k \in Q_w \text{ and } (w,a_k,v) \in \Delta \}$ with
$Q_v = \{k a_k v \mid k a_k v \in K' \}$
and $k a_k v \sim_i k' a_k v' \iff k \sim_i^{\calK} k'$ and $v \sim_i^{\calG} v'$.
By taking the connected components of this model under 
the coarsening $\sim^\cup := \bigcup_{i = 0}^{n-1}\!\!\sim_i$, we obtain the 
set of epistemic successor models. 
When starting from 
the trivial model that consists only of the initial node of the game, 
and successively applying the update, one unravels a tree labelled with 
epistemic models, which corresponds to a two-player game of perfect information
where the strategies of one player translate to coordination
strategies of the grand coalition in the original game, and vice versa. This tree structure, which in general may contain infinitely many distinct labels for its nodes (the undecidable game in \cite{BerwangerKai10}, for example), is called the \textit{tracking} of the game structure.
 
The main result of~\cite{BKP11} shows that, whenever two nodes of the unravelling 
tree carry homomorphically equivalent labels, they can be identified without changing the 
(winning or losing) status of the game. This holds for all imperfect-information games 
with $\omega$-regular winning conditions that are \emph{observable}.
Consequently, the strategy synthesis problem is decidable for a subclass of such games, whenever 
the unravelling process is guaranteed to generate only finitely many epistemic models, 
up to homomorphic equivalence.

Let us now consider the tracking of a game $\calG$ with an observable $\omega$-regular winning condition. 
We claim that every history where the grand coalition attains certainty leads to an epistemic model 
that is homomorphically equivalent to the trivial structure consisting of a singleton labelled with the 
(certain) state at which the history ends. This is because every $\sim^\cup$-connected component is also 
$\sim^0$-connected, and all histories in such a component end at the same state. 
On the other hand, when updating an epistemic model,
the successor models can be at most exponentially larger (for fixed action space).
The property of periodic certainty implied by
recurring certainty, allows us to conclude that the number of updating rounds in which the 
models can grow is bounded by the certainty period of $G$. 
Therefore, games with recurring certainty have finite tracking. 
By~\cite{BKP11}, this implies that the winner determination problem is decidable for such games, 
and finite-state winning strategies can be effectively synthesised whenever the grand coalition 
has a winning strategy.  
    
\bibliographystyle{eptcs}
\bibliography{newall}

\end{document}